\documentclass[useAMS,usenatbib]{mn2e}
\usepackage{epsfig}
\usepackage{amsmath}

\newcommand{\be}{\begin{equation}}
\newcommand{\beq}{\begin{equation}}
\newcommand{\ba}{\begin{eqnarray}}
\newcommand{\ee}{\end{equation}}
\newcommand{\eeq}{\end{equation}}
\newcommand{\ea}{\end{eqnarray}}

\def\lsim{~\rlap{$<$}{\lower 1.0ex\hbox{$\sim$}}}

\def\gsim{~\rlap{$>$}{\lower 1.0ex\hbox{$\sim$}}}

\title[Mg-II absorbers in GRB afterglow spectra]{A gravitational lensing explanation for the excess of strong Mg-II absorbers in GRB afterglow spectra}

\author[Wyithe, Oh \& Pindor]{J. Stuart B. Wyithe$^1$, S. Peng Oh$^2$ and Bartosz Pindor$^1$\\$^1$
School of Physics, University of Melbourne, Parkville, Victoria,
Australia\\$^2$ Department of Physics, University of California, Santa Barbara\\Email: swyithe@physics.unimelb.edu.au}

\begin{document}


\maketitle

\label{firstpage}
\begin{abstract}
GRB afterglows offer a probe of the intergalactic medium out to high redshift which complements observations along more abundant quasar lines-of-sight. Although both quasars and GRB afterglows should provide a-priori random sight-lines through the intervening IGM, it has been observed that strong Mg-II absorbers are twice as likely to be found along sight-lines toward GRBs. Several proposals to reconcile this discrepancy have been put forward, but none has been found sufficient to explain the magnitude of the effect. In this paper we estimate the effect of gravitational lensing by galaxies and their surrounding mass distributions on the statistics of Mg-II absorption. We find that the multi-band magnification bias could be very strong in the spectroscopic GRB afterglow population and that gravitational lensing can explain the discrepancy in density of absorbers, for plausibly steep luminosity functions. The model makes the prediction that approximately 20\%-60\% of the spectroscopic afterglow sample  (i.e. $\sim5-15$ of 26 sources) would have been multiply imaged, and hence 
result in repeating bursts. We show that despite this large lensing fraction it is likely that none would yet have been identified by chance owing to the finite sky coverage of GRB searches. We predict that continued optical monitoring of the bright GRB afterglow locations in the months and years following the initial decay would lead to identification of lensed GRB afterglows.
A confirmation of the lensing hypothesis would allow us to constrain the GRB luminosity function down to otherwise inaccessibly faint levels, with potential consequences for GRB models.

\end{abstract}


\section{Introduction}
   
At moderate redshifts ($0.3\la z\la2.2$) the dispersal of metal enriched gas through the IGM via galactic winds is most readily traced via Mg-II absorption systems in optical spectra of bright  background sources. Although the detailed origin of Mg-II absorbers remains uncertain it is thought that they are associated with galaxies and galactic outflows. For example, Mg-II absorbers have been shown to be associated
with neutral hydrogen absorbers over a range of column densities, including damped Lyman-$\alpha$ absorbers \citep[e.g.][]{rao2006}.  Moreover the host halo masses associated with MgII absorbers have been estimated at $z\sim0.5$ via cross-correlation with luminous red galaxies in the Sloan Digital Sky Survey Data Release 3 \citep[][]{bouche2006}, yielding a host mass of $M\sim10^{12}M_\odot$ (i.e. massive galaxies). Indeed the large number of MgII absorbers and galaxies available for cross-correlation yields a statistical accuracy of a factor of two in host halo mass.  

Obtaining an unbiased census of the density and distribution of Mg-II in the IGM requires that the back-ground sources be uncorrelated with the foreground absorbers under study. Most current knowledge regarding the census of Mg-II absorption systems comes from spectra of quasars \citep[e.g.][]{prochter2006b}. More recently, observations of GRB afterglows have begun to offer a new probe of the intergalactic medium out to high redshift which complements the more abundant quasar lines-of-sight. Indeed, since they are associated with star formation rather than supermassive black holes (which, given their long assembly times, become extremely rare at high redshift), GRBs could potentially be seen out to much higher redshift than quasars; the current record holder is at $z\sim8.1$ \citep{salvaterra09}.   
Interestingly, although both quasars and GRB afterglows should provide a-priori random sight-lines through the intervening IGM it has been observed that strong Mg-II absorbers are several times as likely to be found along sight-lines to GRBs as along quasar sightlines \citep[][]{prochter2006,tejos2009,vergani2009}, indicating that one or both of these samples are 
biased relative to the underlying population of absorbers. Several proposals to reconcile this discrepancy have been put forward, with a detailed discussion of possible biases presented by \citet[][]{porciani2007}. For example, the incidence of  Mg-II systems in quasars could be lowered because of dust obscuration associated with the absorbing systems; 
this turns out to be too small to explain the discrepancy. Alternatively, it has been argued that different sizes of the source could lead to different absorber incidence between GRB afterglows and quasars \citep[][]{frank2007}. However, 
from the similarity of the equivalent width distributions in GRBs and quasars, 
\citet[][]{porciani2007} show that the absorbers must be larger than the beamsize, and hence it is not possible to explain the difference in this way. 

A third potential bias is provided through gravitational lensing 
of GRBs by foreground galaxies associated with the Mg-II absorber. There are several lines of circumstantial evidence for this. Firstly, imaging studies of the host galaxies of GRBs with early time afterglow spectra \citep[][]{chen2009} show that additional galaxies are found within $2''$ of all GRB host fields where the line-of-sight contains a Mg-II absorber, while no additional galaxies are found within $2''$ of GRB lines-of-sight that do not have any Mg-II absorbers. Moreover, GRB afterglows that have more than one absorber are found to be a factor of 1.7 brighter than 
the others \citep[][]{porciani2007}. Under the assumption that gravitational lensing magnification of GRBs is responsible for the excess number of absorbers in GRB spectra relative to quasars, \citet[][]{tejos2009} estimated the 
fraction of GRB afterglows that  need to have been included in the sample owing to magnification. They find a value of 60\% which indicates a large gravitational lensing induced magnification bias. On the other hand, no strongly lensed GRBs have been discovered.\footnote{The expected rate of strongly lensed GRBs was discussed by \citet[][]{porciani2001}, who found that although a few strongly lensed GRBs should be present in a 3 year survey with $Swift$, the partial sky coverage means that it is unlikely that lensed pairs would have been identified as recurrent bursts.}
At first sight this suggests that gravitational lensing could not provide the required explanation for the excess absorbers in GRB afterglow spectra. However, as noted by \citet[][]{porciani2007} the selection of GRB afterglows in two (apparently) independent bands (namely gamma-rays and the optical) leads to the possibility of an increased multi-band magnification bias \citep[][]{wyithe2003}. One goal of this paper is to determine under which circumstances a sufficiently large magnification bias can be obtained.

 The most recent collection of GRB afterglow spectra from which strong ($>1$\AA) Mg-II absorption can be studied is summarised in the work of \citet[][]{vergani2009} who compiled a list of 26 spectra (with a combined redshift path-length of $\Delta z=31.55$) containing 22 strong Mg-II absorbers among 15 of the 26 lines-of-sight.    From that paper we take the following values for observables describing the absorber population. 
The fraction of lines-of-sight that contain one or more absorbers is $F^{\rm obs}_{\rm Mg}\sim0.6\pm0.15$. By contrast, the results of high resolution spectroscopy from \citet[]{prochter2006b} yielded 22 absorbers along 91 quasar lines-of-sight, implying that $F_{\rm Mg,q}=0.25$. Assuming the same redshift pathlength distribution as the GRB sample, these values imply that the relative incidence of the number of strong systems in GRB afterglows relative to quasars is $R^{\rm obs}_{\rm GRB,q}\sim2.1\pm0.6$. Where required we take the redshifts of GRBs and associated absorbers, as well as the probed redshift path-lengths from this paper. Our goal is to make a quantitative comparison of a simple parameterised model for the expected number of absorbers (which includes the bias introduced by gravitational lensing), with these observations.

Our paper is presented as follows. Section~\ref{model} describes the basics of the lensing model with which we interpret the absorber statistics. In \S~\ref{statistics} and \S~\ref{comparison} we discuss the statistics of Mg-II absorbers in GRB afterglow spectra, and their comparison with statistics from quasar lines-of-sight within the context of our model parameters. We then discuss the predicted distribution of absorber redshifts (\S~\ref{absorber})  and observed source magnifications (\S~\ref{magnification}). In \S~\ref{strong} we discuss the rate of multiple imaging among the spectroscopic GRB afterglow sample before presenting our summary in \S~\ref{summary}. In our numerical examples, we adopt the standard set of cosmological parameters \citep[][]{komatsu2009}, with values of $\Omega_{\rm b}=0.04$, $\Omega_{\rm m}=0.24$ and $\Omega_Q=0.76$ for the matter, baryon, and dark energy fractional density respectively, and $h=0.73$, for the dimensionless Hubble constant.
   
\section{lensing model}
\label{model}

The basis of our model is that we assume all lines-of-sight within a distance 
\begin{equation}
R = R_{0}\left(\frac{\sigma}{200\mbox{km/s}}\right)(1+z)^{-1}
\end{equation}
of the nearest galaxy \citep[e.g.][]{tinker2008} contain a Mg-II absorption system with an equivalent width greater than some specified minimum (taken as $>1$\AA\, for the Mg-II absorption statistics modelled in this paper), while lines-of-sight with impact parameters larger than $R$ contain no absorbers. This scaling is chosen to represent a constant multiple of the virial radius. Here $R_{0}$ is the proper radius within which a strong absorber will be found around a $200 $km$\,$s$^{-1}$ halo at $z=0$. We then evaluate the probability of a particular line-of-sight containing a Mg-II absorber, including the effects of magnification bias owing to gravitational lens induced magnification of the observed source flux. For simplicity we model galaxies as singular isothermal spheres (SIS), which is an adequate approximation for the purposes of this paper \citep[][]{treu2004}. In the case of close alignment, the presence of an intervening galaxy can produce strongly lensed multiple images of background sources, which are magnified relative to their intrinsic flux (with a magnification $\mu>2$ for the bright image in the case of an SIS).  Images that are not so closely aligned, and therefore only singly imaged can also be magnified (in the case of an SIS by a value of $1<\mu<2$). In this paper we distinguish between the numbers of absorbers in the spectra of multiply imaged and singly imaged sources. 

We refer to the a-priori probability for a source to be multiply imaged as the multiple image optical depth $\tau_{\rm m}$. In addition to this quantity we calculate the optical depth $\tau_{\rm Mg}$ for Mg-II absorption (i.e. the probability that a line-of-sight goes within $R$ of a galaxy), as well as the optical depth for which there are no absorbers $\tau_{\rm noMg}=1-\tau_{\rm Mg}$. The optical depth must be computed over the redshift path-lengths where Mg-II absorption has been searched. In the case of GRB afterglow spectra this is object dependent since different GRBs were observed with a range of telescopes and spectral ranges. 
Accounting for this, the multiple image and Mg-II absorption optical depths are
\begin{equation}
\tau_{\rm m}=\frac{1}{N_{\rm GRB}}\sum_{i=1}^{N_{\rm GRB}} \int_{z_{{\rm l},i}}^{z_{{\rm h},i}} dz  \int d\sigma \Phi(\sigma) (1+z)^3 \frac{cdt}{dz} \pi D_{\rm d}^2 \theta_{\rm ER}^2(\sigma) 
\end{equation}
and 
\begin{equation}
\tau_{\rm Mg}=\frac{1}{N_{\rm GRB}}\sum_{i=1}^{N_{\rm GRB}} \int_{z_{{\rm l},i}}^{z_{{\rm h},i}} dz  \int d\sigma \Phi(\sigma) (1+z)^3 \frac{cdt}{dz} \pi R^2(\sigma)
\end{equation}
respectively, where $\theta_{\rm ER}$ is the Einstein radius and $D_{\rm d}$ is the angular diameter distance to the lens. 
Here we assume that the SDSS velocity function \citep[][]{sheth2003} describes the galaxy population (and assume a constant co-moving density at fixed $\sigma$)

\begin{equation}
\Phi(\sigma) d\sigma = \Phi_\star\left(\frac{\sigma}{\sigma_\star}\right)\frac{\exp{[-(\sigma/\sigma_\star)^\beta]}}{\Gamma(\alpha/\beta)}\beta\frac{d\sigma}{\sigma},
\end{equation}
where $\Phi_\star=2\times10^{-3}$Mpc$^{-3}$, $\alpha=6.5$, $\beta=14.75/\alpha$ and $\sigma_\star=161\Gamma(\alpha/\beta)/\Gamma((\alpha+1)/\beta)$. To calculate $\tau_{\rm m}$ we use the expression for the angular Einstein radius for an SIS
\begin{equation}
\theta_{\rm ER}(\sigma)=1.4'\frac{D_{\rm ds}}{D_{\rm s}}\left(\frac{\sigma}{200\mbox{km/s}}\right)^2,
\end{equation}
where $D_{\rm s}$ and $D_{\rm ds}$ are the angular diameter distances to the source, and between the lens and source respectively.

The GRB afterglow sample is selected from the wider GRB population using two flux limits. Firstly there is a selection in gamma-rays, and secondly a selection for optical afterglows that are bright enough to facilitate the acquisition of a spectrum with sufficient signal-to noise.  Through magnification bias, these flux limits lead to a bias on the probability with which GRB afterglows enter the sample as a function of magnification, and hence to a bias of impact parameter relative to the nearest galaxy. The result is that magnification bias concentrates observed sources\footnote{Gravitational lensing also lowers the density by magnifying the angular extent of the image plane relative to the source plane. This effect, which is usually referred to as depletion, is not dominant for steep luminosity functions.} in a flux limited sample around foreground objects \citep[][]{webster1988}, which in the context of Mg-II absorbers associated with galaxies should lead to an excess of absorbers per unit redshift in a gravitationally lensed sample. 

To quantify the potential magnitude of the multi-band magnification bias we specify two cumulative luminosity functions $\Psi_{\gamma}(L_{\gamma})$ and $\Psi_{\rm A}(L_{\rm A})$ describing the number densities of sources above fixed apparent 
gamma-ray $L_{\gamma}$ and optical $L_{\rm A}$ luminosities. With these the bias for observing a lensed source relative to the unlensed population is \citep[][]{wyithe2003}
\begin{equation}
B = \frac{1}{\Psi_{\gamma}(L_{\gamma})\Psi_{\rm A}(L_{\rm A})}\int_{\mu_{\rm min}}^{\mu_{\rm max}}d\mu \frac{dP}{d\mu} \Psi_{\gamma}(L_{\gamma}/\mu)\Psi_{\rm A}(L_{\rm A}/\mu)
\end{equation}
where $dP/d\mu$ is the probability distribution for magnification within the range $\mu_{\rm min}<\mu<\mu_{\rm max}$. We then calculate the bias for the singly imaged Mg-II absorbers using the distribution,
\begin{equation}
\frac{dP_{\rm Mg}}{d\mu}=2\frac{(\mu-1)^{-3}}{\left((\mu_R-1)^{-2}-1\right)^{-1}}\hspace{5mm}\mbox{for}\hspace{3mm}\mu_R<\mu<2,
\end{equation}
where $\mu_R=1+\theta_{\rm ER}D_{\rm d}/R$,
as well as the bias for each of the two multiple images using the distributions
\begin{equation}
\frac{dP_{{\rm m},1}}{d\mu} = \frac{2}{(\mu-1)^3}    \hspace{5mm}\mbox{for}\hspace{3mm} 2<\mu<\infty
\end{equation}
and
\begin{equation}
\frac{dP_{{\rm m},2}}{d\mu} = \frac{2}{(\mu+1)^3}    \hspace{5mm}\mbox{for}\hspace{3mm} 0<\mu<\infty.
\end{equation}
Note that we treat the two images of a multiply imaged GRB separately since they have a time delay which is typically longer than the duration of the afterglow decay time. To conserve flux we approximate the magnification of regions with no Mg-II absorber using a constant magnification ($\mu_{\rm noMg}<1$) which normalises the mean of the overall magnification distribution to unity.

\begin{figure*}
\includegraphics[width=18cm]{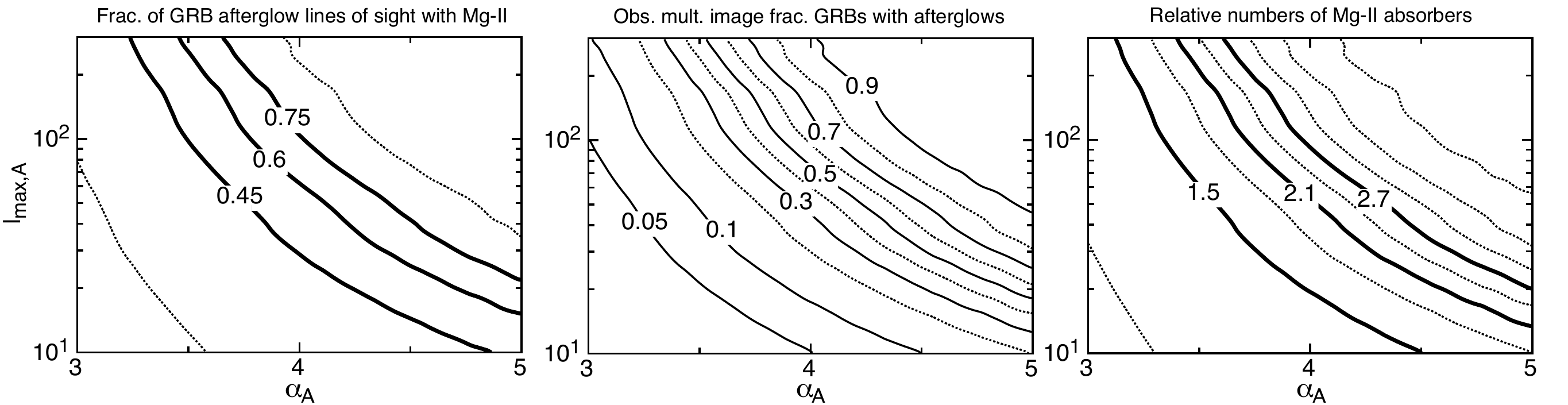}
\caption{Panels showing contours over the parameter space of $\alpha_{\rm A}$ and $l_{\rm max}$ for $i)$ the fraction of GRB lines-of-sight with Mg-II absorbers ($left$), $ii)$ the fraction of multiply imaged GRBs among the afterglow sample ($centre$) and $iii)$ the ratio of absorbers per unit redshift predicted for GRB afterglow and quasar spectra ($right$). The model assumed a value of $R_{0}=165$kpc, for which the fraction of lines-of-sight which would contain a Mg-II absorber in the absence of lensing is 0.25. In the left and right panels, the heavy contours correspond to the mean and 1-sigma uncertainties on the quantities $F_{\rm Mg}$ and $R_{\rm GRB,q}$ respectively.}
\label{fig1}
\end{figure*}

By combining the above expressions for the optical depth and magnification bias we next calculate the corresponding  biased optical depths for spectra that do not contain a Mg-II absorber 
\begin{equation}
\mathcal{B}_{\rm noMg}=B_{\rm noMg}\tau_{\rm noMg},
\end{equation}
as well as the biased optical depths for spectra that contain a Mg-II absorber within a spectrum of a singly imaged source, and within the spectra of a multiply imaged source
\begin{eqnarray}
\nonumber
&&\mathcal{B}_{\rm s}=\int_0^\infty d\sigma\int_0^\infty dz \frac{d^2\tau_{\rm Mg}}{d\sigma\,dz} B_{\rm Mg}(z,\sigma)\tau_{\rm Mg}(z,\sigma)\\
\nonumber
&&\mathcal{B}_{{\rm m},1}=B_{{\rm m},1}\tau_{\rm Mg}\\
&&\mathcal{B}_{{\rm m},2}=B_{{\rm m},2}\tau_{\rm Mg}
\end{eqnarray}
In the above $\mathcal{B}_{\rm s}$ involves integrating the differential optical depth over redshift and velocity dispersion because the bias is dependent on these quantities, whereas for multiple imaging the magnification probability distribution (and hence bias) are independent of these quantities. 
From these biased optical depths we are able to obtain the corresponding fractions of observed GRB afterglow spectra that do not contain an Mg-II absorber 
\begin{equation}
F_{\rm noMg} = \frac{\mathcal{B}_{\rm noMg}}{\mathcal{B}_{\rm noMg} + \mathcal{B}_{\rm s} + \mathcal{B}_{{\rm m},1} + \mathcal{B}_{{\rm m},2}}
\end{equation}
as well as the fractions that contain a Mg-II absorber within a spectrum of a singly or multiply imaged GRB afterglow,
\begin{eqnarray}
F_{\rm s} &=& \frac{\mathcal{B}_{\rm s}}{\mathcal{B}_{\rm noMg} + \mathcal{B}_{\rm s} + \mathcal{B}_{{\rm m},1} + \mathcal{B}_{{\rm m},2}},\\
F_{\rm m} &=& \frac{\mathcal{B}_{{\rm m},1} + \mathcal{B}_{{\rm m},2}}{\mathcal{B}_{\rm noMg} + \mathcal{B}_{\rm s} + \mathcal{B}_{{\rm m},1} + \mathcal{B}_{{\rm m},2}}.
\end{eqnarray}
The total fraction of GRB lines of sight which contain a MgII absorber is $F_{\rm Mg}=F_{\rm s} + F_{\rm m}$. 
In the following section we use the above model to calculate the fraction of lines-of-sight with Mg-II absorbers among the GRB afterglow sample. 

\section{Statistics of absorbers in GRB afterglows} 
\label{statistics}

To calculate the fraction of GRB lines-of-sight with Mg-II absorbers we must specify luminosity function models for both the GRBs and their optical afterglows. The gamma-ray and optical luminosities are thought to be independent 
\citep{nardini06,nardini2008}, and so we specify the luminosity functions in these bands separately. For the GRBs we assume a shallow power-law of the form $\Psi_{\gamma}(L_{\gamma})\propto (L_\gamma)^{-\alpha_{\gamma}}$, and set $\alpha_{\gamma}=0.7$ to correspond to the measured differential slope of $\sim1.7$ \citep[][]{schaefer01}.
The luminosity function of optical afterglows is much less well-constrained. The observed distribution appears bimodal and relatively narrow, especially for the high luminosity branch \citep{nardini2008}. Furthermore, many ($\sim 40\%$ of long GRBs) are not observed to have optical counterparts. The latter fact suggests that the slope of the optical afterglow luminosity function could be steep. 
Therefore, for the optical afterglows we also assume a power-law $\Psi_{\rm A}(L_{\rm A})\propto (L_{\rm A})^{-\alpha_{\rm A}}$, but taking into account the possibility that the slope is very steep. We note that for cumulative slopes steeper than $\alpha_{\rm A}>2$ the magnification bias becomes formally infinite. 
This would result in an unobserved plethora of multiply-imaged GRBs. 
 We therefore introduce a break in the luminosity function below which the slope is assumed to be very flat so that intrinsically faint objects do not contribute to the magnification bias. Physically, this is also necessary to ensure that the number density of afterglows does not diverge. 
If we assume there are no sources fainter than $L_{\rm break}$, 
we can characterize the magnification required to view sources with the break luminosity $L_{\rm break}$ as
$\mu_{\rm max}=l_{\rm max}\equiv L_{\rm lim,A}/L_{\rm break}$.  Thus given an observed luminosity limit $L_{{\rm lim,A}}$ the 
intrinsic multi-band luminosity function is assumed to be the product of the individual single-band luminosity functions
\begin{eqnarray}
\nonumber
&&\hspace{-7mm}\Psi_{\gamma,{\rm A}}(L_{\gamma},L_{\rm A})\propto (L_{\gamma})^{-\alpha_{\gamma}}\left(\frac{L_{\rm A}}{L_{{\rm break}}}\right)^{-\alpha_{\rm A}}\hspace{0mm} \mbox{where}\hspace{1mm}L_{\rm A}>L_{\rm break}\\
\nonumber
&&\hspace{-7mm}\Psi_{\gamma,{\rm A}}(L_{\gamma},L_{\rm A})\propto (L_{\gamma})^{-\alpha_{\gamma}}\hspace{21mm} \mbox{where}\hspace{1mm}L_{\rm A}<L_{\rm break}\\
\end{eqnarray}

\subsection{Mg-II absorbers in quasar lines-of-sight}
In order to compare the statistics of GRB and quasar Mg-II absorbers we must also compute the fraction of quasar lines-of-sight that contain a Mg-II absorber ($F_{\rm Mg,q}$). This modelling is based on the model of GRB afterglow statistics described above, with the following modifications. Firstly, since quasars are selected in correlated (optical) bands there is only a single factor of magnification bias. Moreover we assume that the flux limit in these samples corresponds to quasars that are at the faint end of the luminosity function, as is appropriate for large modern quasar surveys such as the SDSS \citep[we assume a cumulative slope of $\alpha_{\rm q}=0.4;$][]{croom2009}. 
For such a shallow slope, the magnification bias is negligible; single-band magnification bias only becomes important for a cumulative slope of $\alpha_{\rm q} \gsim 2$. 
In addition, since quasars are not transient, the two images of a strongly lensed source are considered together rather than as separate sources.  We note as a consistency check that our model predicts a multiply imaged lensed fraction of $0.001$ for quasars at $z\sim2$ which is consistent with previous work \citep[e.g. ][]{pindor2003}. 

\begin{figure*}
\begin{center}
\includegraphics[width=13cm]{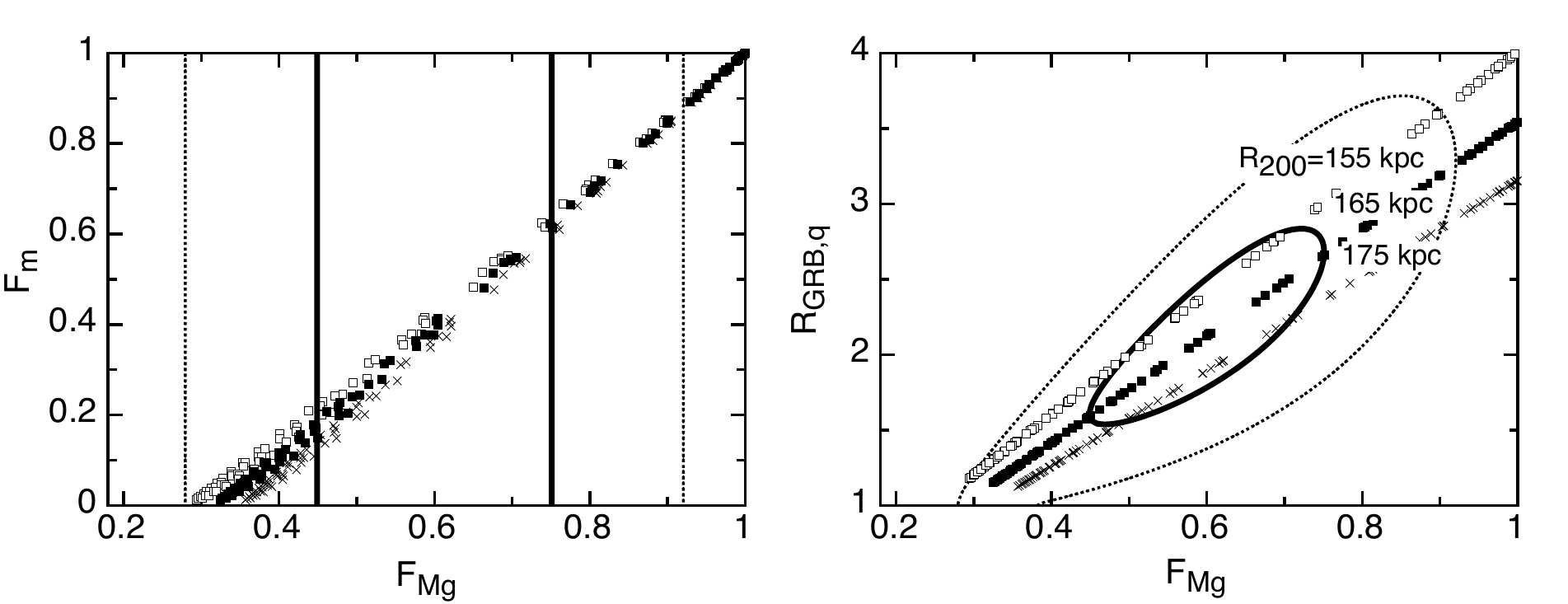}
\end{center}
\caption{Scatter plots of points in the model parameter space $(\alpha_{\rm A},l_{\rm max})$. In the $left$ panel we show the relation between the fraction of GRB lines-of-sight with Mg-II absorbers, and the fraction of multiply imaged GRBs. In the $right$ panel we show the relationship between the fraction of GRB lines-of-sight with Mg-II absorbers, and the ratio of absorbers per line-of-sight between GRBs and quasars. In each case we show  show the relationships assuming 3 values of $R_{0}$ in order to show the sensitivity to this parameter. To show the correspondence of the model to observed absorber statistics we have plotted in each panel contours at 60\% and 10\% of the peak likelihood given the constraints $F_{\rm Mg}^{\rm obs}=0.6\pm0.15$ and $F_{\rm Mg,q}^{\rm obs}=0.28\pm0.05$.}
\label{fig2}
\end{figure*}

\subsection{Results}
To produce a realisation of our model we therefore need to specify three parameters. Firstly there is the value of $R_{0}$, which in our fiducial case we set to $165$kpc. This value reproduces the observed $F_{\rm Mg,q}^{\rm obs}=0.28$ for the fraction of quasar lines-of-sight with a strong Mg-II absorber \citep[and is consistent with the high resolution spectroscopy of ][which yielded 22 absorbers along 91 quasar lines-of-sight]{prochter2006b}.
We then compute models for a range of values of the very uncertain parameters describing the GRB afterglow LF, $\alpha_{\rm A}$ and $l_{\rm max}$. The results are presented in the left and central panels of Figure~\ref{fig1}, where contours of the fraction of GRB lines-of-sight with Mg-II absorbers, and the fraction of multiply imaged GRBs among the afterglow sample are shown respectively. As expected there is clear degeneracy between the parameters $l_{\rm max}$ and $\alpha_{\rm A}$ since large values of each parameter lead to large magnification biases and hence to large fractions of lines-of-sight with Mg-II absorbers as well as large multiple image fractions. 

To more clearly understand the relationship between the fraction of GRB lines-of-sight with Mg-II absorbers, and the fraction of multiply imaged GRBs among the afterglow sample we plot their corresponding values for all points in the model parameter space $(\alpha_{\rm A},l_{\rm max})$ in the left panel of Figure~\ref{fig2}. Figure~\ref{fig2} demonstrates the small amount of uncertainty introduced into this relationship through the unknown parameters $(\alpha_{\rm A},l_{\rm max})$, and indicates that larger fractions of lines-of-sight with Mg-II absorption correspond to larger multiple image fractions. 
This strong correlation is unsurprising, since $F_{\rm Mg}=F_{\rm s} + F_{\rm m}$. However, note that both $F_{\rm s}$ and $F_{\rm m}$ depend on the lens model. In particular, $F_{\rm s}$ falls as lensing becomes more important since if either $F_{\rm Mg} \rightarrow 1$, $F_{\rm m} \rightarrow 1$, or $F_{\rm s} \rightarrow 0$ the  unlensed or singly imaged lines of sight become negligible even though $F_{\rm Mg}=F_{s}=F_{\rm Mg,q}\approx 0.3$ in the absence of lensing.   
In Figure~\ref{fig2} we also show corresponding sets of points for 2 additional values of $R_{0}$ lying within the range of uncertainty governed by the observed quantity $F_{\rm Mg,q}^{\rm obs}=0.28\pm0.05$ \citep[where the uncertainty is estimated from the number of strong absorbers in the high resolution spectroscopic sample of][]{prochter2006b} in order to show the sensitivity to this parameter. Given the allowed values of $R_{0}$, we can read off the predicted multiple image fraction from the upper panel of Figure~\ref{fig2}, which shows that having Mg-II absorbers in $\sim(60\pm15)\%$ of GRB afterglow lines-of-sight indicates a multiply lensed fraction of $F_{\rm m}\sim20\%-60\%$.

\section{Comparison of absorber statistics along quasar and GRB afterglow lines-of-sight} 
\label{comparison}

As alluded to in the introduction, a hitherto unresolved puzzle has been the fact that quasar lines-of-sight show a factor of $R^{\rm obs}_{\rm GRB,q}\sim2.1\pm0.6$ fewer Mg-II absorbers per unit redshift than do GRB afterglow lines-of-sight. Gravitational lensing has been proposed as a possible solution to this puzzle \citep[][]{porciani2007,tejos2009}, though no quantitative calculation has been provided. In this section we use our gravitational lensing model to determine whether gravitational lensing can explain the difference in absorber density, and if so, for what regions of parameter space describing the unknown GRB afterglow luminosity function. 

In the right hand panel of  Figure~\ref{fig1} we show contours of the ratio of absorbers per unit redshift predicted for GRB afterglow and quasar spectra in our model ($R_{\rm GRB,q}=F_{\rm Mg}/F_{\rm Mg,q}$) over the parameter space of $\alpha_{\rm A}$ and $l_{\rm max}$. We find that ratios of $R_{\rm GRB,q}\sim2$ can be obtained for cumulative slopes of $\alpha_{\rm A}\sim3.5$ or steeper provided the power-law extends sufficiently far below the flux limit. As before there is degeneracy and to more clearly understand the relationship between the fraction of GRB lines-of-sight with Mg-II absorbers and the ratio of absorbers per line-of-sight between GRBs and quasars we plot their corresponding values for all points in the model parameter space $(\alpha_{\rm A},l_{\rm max})$ in the right panel of figure~\ref{fig2}. As before we have plotted this relation for three values of $R_{0}$.

\section{Absorber redshift distribution}
\label{absorber}

 \begin{figure*}
\includegraphics[width=18cm]{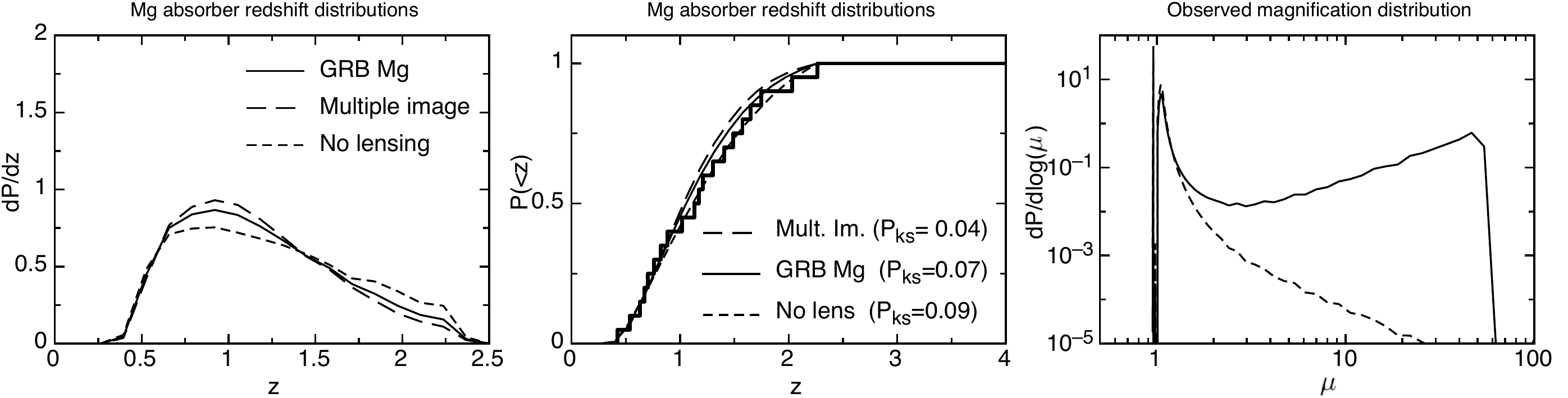}
\caption{
The $left$ and $central$ panels show differential and cumulative absorber redshift distributions respectively. We have picked a fiducial model of $\alpha_{\rm A}=4$ and $l_{\rm max}=50$ for this example. Three model distribution are shown corresponding to our absorber model (solid lines), to the distribution of gravitational lens redshifts (long dashed lines), and a distribution that assumes no lensing. The cumulative distributions (central panel) are compared to the observed absorber redshift distribution (thick stepped line) and the probabilities associated with a KS test between the models and data are listed. The observed (solid line) and intrinsic (dashed line) magnification distributions for the fiducial model is shown in the $right$ hand panel.}
\label{fig3}
\end{figure*}

In addition to the statistics of the number of absorbers relative to quasars, it is important that the model also successfully reproduce the observed absorber redshift distribution. The predicted differential and cumulative redshift distributions for a fiducial model ($R_{0}=165$kpc) of $\alpha_{\rm A}=4$ and $l_{\rm max}=50$, giving $R_{\rm GRB,q}\sim2.4$  and $F_{\rm m}\sim0.3$,
are shown in the left and central panels of Figure~\ref{fig3} respectively. Three model distributions are shown, with the absorber redshift distribution strongly constrained a-priori by the path-lengths probed by the spectra in each case.  The solid curves represent the absorber redshift distribution predicted by our model. 

In addition to the Mg-II absorber model, we show for comparison the distribution of gravitational (multiple image) lens redshifts 
(long dashed lines). This distribution is more peaked than predicted by the absorber model which includes both lensed and unlensed sources. We also show a model that assumes no lensing, which leads to a distribution that is more spread out in redshift. The cumulative distributions (central panel) are compared to the observed absorber redshift distribution (thick stepped line). We have performed a KS test between each of the three model distributions and the observed distribution. The probabilities of finding a distribution of absorber redshifts drawn from the model that are less consistent with the parent distribution than the observations are listed. 
While the data is unable to strongly distinguish between the different models, the hypothesis of all Mg-II absorbers lying along strongly lensed lines of sight is somewhat less consistent ($P_{\rm ks}=0.04$) than the mixed absorber or no-lens hypothesis ($P_{\rm ks}=0.07,0.09$ respectively). 

\section{Observed magnification distribution}
\label{magnification}

In addition to modifying the fraction of sources in a flux limited sample that are seen as multiple images, gravitational lensing magnification bias also modifies the distribution of source magnifications relative to the intrinsic magnification distribution. The predicted  distribution of observed magnification for the fiducial model ($\alpha_{\rm A}=4$, $l_{\rm max}=50$) is shown in the right hand panel of Figure~\ref{fig3}. There are 3 features of note in this distribution. The two peaks near magnifications of unity result from our assumption that a magnification within $R$ of a galaxy is given by the SIS, while beyond $R$ the magnifications all have a constant value (leading to the delta function seen in the figure).  A more sophisticated analysis would draw on a global distribution for lensing by large scale structure \citep[e.g.][]{fluke2002}, however we note that since magnifications near unity lead to negligible magnification bias, this does not affect the results of our paper. 

In addition to these low magnification peaks, our model also predicts a prominent additional peak at high magnification. This peak differs from the generic shape of $dP/d\mu\propto \mu^{-3}$ for strong lenses owing to magnification bias which greatly increases the number of sources present in the flux limited sample from regions of high magnification. 
As $\alpha_{A},l_{\rm max}$ increase, the number of such highly magnified sources increase as well. 
The sharp cutoff in the distribution beyond $\mu=50$ corresponds to our assumed value for $l_{\rm max}$. It is interesting to note that the GRB afterglow luminosity function is thought to be bi-modal \citep[][]{nardini2008}. Naively, the bi-modality of the magnification distribution suggests a possible lensing origin for bimodality in the observed GRB afterglow luminosity function. However when the intrinsic magnification distribution is convolved with the luminosity function it is easy to show that lensing cannot produce a bimodal observed luminosity function where the intrinsic luminosity function is monotonically decreasing with luminosity (instead, a large magnification bias leads to a slope of $\alpha_{\rm A}=-2$ above the flux limit). We note that the observed bi-modal distribution is not inconsistent with the lensing hypothesis, because the steep power-law need only be present below the flux limit in order to produce a large magnification bias.

\section{The number of strongly lensed GRB afterglows}
\label{strong}

One striking prediction from our model is that if gravitational lensing is responsible for the observed difference in the number of Mg-II absorbers per unit redshift in GRB afterglow and quasar spectra, then of order 30\% ($\sim20\%-60\%$) of the GRBs with afterglow follow up should have been multiply imaged. This figure is consistent with the estimate of 60\% from \citet[][]{tejos2009} for the number of GRBs that must have been included in the sample owing to magnification in order to have produced the observed $R_{\rm GRB,q}^{\rm obs}=2-3$, and implies that $\sim9$ ($\sim5-15$) of the current sample of 26 spectroscopic afterglows were multiply imaged. On the other hand no multiply imaged GRBs have been discovered, either within the overall GRB sample, or within the afterglow subsample. In this section we therefore discuss whether current observational campaigns would have missed this high multiple image fraction among the spectroscopic afterglow subsample.
 
Unlike multiply imaged quasars, multiple images of GRBs or their afterglows would be seen as repeating bursts since the timescale of variability is shorter than the typical time delay between multiple images. As a result, for serendipitous identification of a multiply imaged GRB, the images would need to be discovered separately. In particular, assuming that the location of a GRB is not monitored beyond the time when the afterglow is no longer visible, the identification of a repeating image of a lensed source would be via a GRB trigger. The relevant statistical question is the following: in the case where a GRB is strongly lensed (i.e. multiply imaged) with one image having been located, in what fraction of cases would a second image also be detected at a later time? Now a follow up image can only be detected in cases where it is the leading image that is discovered first. The fraction of detected images of lensed sources that are leading images is $f_{\rm lead}\equiv\mathcal{B}_{{\rm m},1}/(\mathcal{B}_{{\rm m},1}+\mathcal{B}_{{\rm m},2})$. For our fiducial model we find $f_{\rm lead}=0.56$, which is 
slightly greater than 1/2 owing to the 
increased effects of magnification bias for the leading image. Assuming monitoring coverage that is random with respect to the unknown time delay and that a GRB monitor surveys $\Omega_{\rm GRB}$ sr at any one time, there is a $\Omega_{\rm GRB}/4\pi$ chance that the telescope would be monitoring the region of sky where the lead image of the GRB was identified at this later time. The probability that a particular multiply imaged GRB would be identified is therefore $P_{\rm m}=   f_{\rm lead}\Omega_{\rm GRB}/4\pi\sim0.09$. Given a sample of $N_{\rm GRB}=26$ afterglow spectra, with a $F_{\rm m}=30\%$ lens fraction, the probability of identifying $n$ lenses among an intrinsic sample of $N_{\rm lens}=F_{\rm m}N_{\rm GRB}$ is given by the binomial theorem 
\begin{equation}
P(n)= \frac{N_{\rm lens}!}{n!(N_{\rm lens}-n)!}(P_{\rm m})^{n}(1-P_{\rm m})^{N_{\rm lens}-n}.
\end{equation}
For the Swift telescope, which monitors $\Omega_{\rm GRB}\sim2$ sr of the sky the probability of finding no lenses is therefore $P(0)=47\%$, while the probabilities of having found 1, 2 or 3 lenses are $P(1)=37\%$, $P(2)=13\%$ and $P(3)=3\%$. Thus we find it unsurprising that lensed GRBs among this sample have not been identified as repeating bursts, even prior to considerations regarding the difficulty of their identification. For instance, a search by \citet{veres09} for lensed bursts in the Fermi GRB data found 3 candidates with plausible timescales, but in the end no definitive lensing detections. 

The above considerations apply to random surveys of the GRB sky. On the other hand, the very large predicted lens fraction implies that targeted optical monitoring of the GRBs with multiple MgII absorption in their afterglows
in the months and years following the decay of the initial afterglow light curve would be a very efficient way of identifying these strongly lensed GRBs. 
This would provide a crucial test of our model. 
Lack of identification of lensed afterglows among this sample would indicate that gravitational lensing is unable to explain the difference in absorber statistics between quasars and afterglow lines-of-sight, and therefore that the afterglow LF is not steep ($\alpha_{\rm A}\ga3$) below the spectroscopic flux limit. 
 
Finally it is worth noting that in contrast to strong MgII systems with $W_{\rm r} \ge 1\, {\rm \AA}$, weak MgII systems with $0.07-0.3 \, {\rm \AA} \le W_{\rm r} \le 1\, {\rm \AA}$ have comparable incidence rates towards GRB and QSO lines of sight \citep{tejos2009,vergani2009}. The incidence of CIV also appears to be similar along GRB and QSO lines of sight \citep{tejos07}. This agrees well with our model under the reasonable hypothesis that CIV and weaker MgII absorbers are associated with lower density gas which lie further away from galaxies, and as such are not subject to lensing bias.   

\section{Summary}
\label{summary}

In recent years the rapid optical imaging and spectroscopic followup of GRBs and their afterglows has began to offer a new probe of the intergalactic medium out to high redshift which complements the more abundant quasar lines-of-sight. Studies of the incidence of Mg-II absorption systems along these lines-of-sight have revealed that although both quasars and GRB afterglows should provide a-priori random sight-lines through the intervening IGM, strong Mg-II absorbers are several times as likely to be found along sight-lines to GRBs \citep[][]{prochter2006,tejos2009,vergani2009}. Several proposals to reconcile this discrepancy have been put forward \citep[see e.g.][]{porciani2007}, but none have been quantitatively successful. In this paper we have described a simple model that associates Mg-II absorbers with the foreground galaxy population, and includes a 2-band luminosity function to describe the number densities of GRBs and their afterglows. We have used this model to estimate the effect of gravitational lensing by galaxies and their surrounding mass distributions on the statistics of Mg-II absorption. 

Our model leads to
 two main findings. Firstly, we show that the multi-band magnification bias could be very strong in the GRB afterglow population. Gravitational lensing can explain the discrepancy in the incidence of absorbers between quasar and GRB lines-of-sight for GRB afterglow luminosity functions with cumulative slopes $\alpha_{\rm A}\ga3.5$. Secondly our model makes the prediction that approximately 20\%-60\% (i.e. between $\sim5$ and 15) of the existing afterglow sample would have been multiply imaged, and hence repeating sources. We show that despite this large lensing fraction it is likely that none would yet have been identified by chance owing to the finite sky coverage of GRB searches. As a result we predict that continued optical monitoring of the bright GRB afterglow locations in the months and years following the initial decay would lead to identification of lensed GRB afterglows. A confirmation of the lensing hypothesis would allow us to constrain the GRB luminosity function down to otherwise inaccessibly faint levels, with potential consequences for GRB models.

{\bf Acknowledgments} The research was supported by the Australian Research
Council (JSBW). SPO was supported by NSF grant AST 0908480. JSBW and BP thank the Physics department at UCSB for hospitality during this work. SPO thanks the UCSB Wednesday 'Gastrophysics' group for lively discussions. 

\newcommand{\noopsort}[1]{}

\bibliographystyle{mn2e}
\bibliography{text}

\label{lastpage}
\end{document}